\documentclass[11pt]{article}
\usepackage[textwidth=15.2cm,textheight=22cm]{geometry}
\usepackage{amsmath,amssymb}
\usepackage{latexsym}
\usepackage{multicol}
\usepackage{graphicx}
\usepackage{bm}

\usepackage{tikz}
\usetikzlibrary{trees}
\usepackage{siunitx}
\usepackage{varwidth}
\usepackage{pifont}
\usepackage{cancel}
\newcommand{\xdownarrow}[1]{%
  {\left\downarrow\vbox to #1{}\right.\kern-\nulldelimiterspace}
}
\usetikzlibrary{shapes.geometric, arrows, shadows}

\tikzstyle{forces} = [rectangle, rounded corners, minimum width=2cm, minimum height=0.8cm,text centered, draw=black]
\tikzstyle{spin} = [rectangle, rounded corners, minimum width=1.3cm, minimum height=0.8cm,text centered, draw=black]
\tikzstyle{theory} = [rectangle, rounded corners, minimum width=2cm, minimum height=0.8cm,text centered, draw=gray]
\tikzstyle{arrow} = [thick,->,>=stealth]
\tikzstyle{arrow1} = [thick,<->,>=stealth]
\tikzstyle{line} = [draw, -latex']

\allowdisplaybreaks[1]

\newcommand{\del}{\partial}
\newcommand{\be}{\begin{equation}}
\newcommand{\ee}{\end{equation}}
\newcommand{\ba}{\begin{eqnarray}}
\newcommand{\ea}{\end{eqnarray}}

\newcommand{\rom}[1]{\uppercase\expandafter{\romannumeral #1\relax}}

\newcommand{\E}{E_{7(7)}}

\def\ba{\bar A}

\def\beq{\begin{equation}}
\def\eeq{\end{equation}}

\newcommand{\nn}{\nonumber}

\newcommand{\ndt}{\noindent}

\newcommand{\delp}{{\partial^+}}

\def\bea{\begin{eqnarray}}
\def\eea{\end{eqnarray}}
\def\beas{\begin{eqnarray*}}
\def\eeas{\end{eqnarray*}}
\def\sla{\raise.15ex\hbox{$/$}\kern-.57em}

\def\spa#1.#2{\left\langle#1\,#2\right\rangle}
\def\spb#1.#2{\left[#1\,#2\right]}
\begin{document}

\begin{titlepage}
\begin{flushright}    
{\small $\,$}
\end{flushright}
\vskip 1cm
\centerline{\Large{\bf{Maximal supergravity and the quest for finiteness}}}
\vskip 2cm
\centerline{Sudarshan Ananth$^\dagger$\footnote{Corresponding author, ananth@iiserpune.ac.in}, Lars Brink$^*$\footnote{lars.brink@chalmers.se} and Sucheta Majumdar$^\dagger$\footnote{sucheta.majumdar@students.iiserpune.ac.in}}
\vskip .7cm
\centerline{$\dagger$\,\it {Indian Institute of Science Education and Research}}
\centerline{\it {Pune 411008, India}}
\vskip 0.7cm
\centerline{$^*\,$\it {Department of  Physics, Chalmers University of Technology}}
\centerline{\it {S-41296 G\"oteborg, Sweden}}
\vskip 0.1cm
\centerline{\it {and}}
\vskip 0.1cm
\centerline{\it{Division of Physics and Applied Physics, School of Physical and Mathematical Sciences}}
\centerline{\it{Nanyang Technological University, Singapore 637371}}
\vskip 1cm
\centerline{March 28, 2018}
\vskip 1.5cm
\centerline{\bf {Abstract}}
\vskip 0.5cm
We show that $\mathcal N=8$ supergravity may possess an even larger symmetry than previously believed. Such an enhanced symmetry is needed to explain why this theory of gravity exhibits ultraviolet behavior reminiscent of the finite $\mathcal N=4$ Yang-Mills theory. We describe a series of three steps that leads us to this result.
\vskip 4cm
\begin{center}
Essay written for the Gravity Research Foundation 2018 Awards for Essays on Gravitation
\end{center}
\vfill
\end{titlepage}

\newpage
\ndt Quantum Field Theory describes three of the four fundamental forces in Nature with great precision. However, when attempts have been made to use it to describe the force of gravity, the resulting field theories, are without exception, ultraviolet divergent and non-renormalizable. One striking aspect of supersymmetry is that it greatly reduces the divergent nature of quantum field theories. Accordingly, supergravity theories have less severe ultraviolet divergences.
\vskip 0.3cm
\ndt Maximal supergravity in four dimensions, $\mathcal N=8$ supergravity~\cite{CJ}, has the best ultraviolet properties of any field theory of gravity with two derivative couplings. Much of this can be traced back to its three symmetries: Poincar\'e symmetry, maximal supersymmetry and an exceptional $E_{7(7)}$ symmetry. In the 1980s, arguments based on power-counting clearly predicted three-loop divergences in this theory. However, explicit four-graviton scattering amplitude computations prove that this theory is ultraviolet finite up to four-loop order~\cite{ZB}. The enhanced cancelations in these calculations render the ultraviolet behavior of $\mathcal N=8$ supergravity very similar to that of $\mathcal N=4$ superYang-Mills theory~\cite{LB}. This is a puzzling result given that the $\mathcal N=4$ theory is ultraviolet finite~\cite{LB2} and conformally-invariant to boot.
\vskip 0.3cm
\ndt Supersymmetry, by itself, cannot explain these magical ultraviolet properties in $\mathcal N=8$ supergravity\footnote{An intriguing possibility is that some of this improved behavior stems from pure gravity itself.}. The onus is therefore placed on the exceptional symmetry in the theory. $\mathcal N=8$ supergravity exhibits a non-linear $E_{7(7)}$ symmetry which is insufficient to account for {\it {all}} the improved properties. Thus our primary motivation is to ask:
\vskip 0.3cm
\ndt {\it {Is there a hidden or enhanced symmetry in $\mathcal N=8$ supergravity that explains its improved ultraviolet behavior and that ultimately proves responsible {\it {if}} this theory is finite to all orders in perturbation theory.}}
\vskip 0.3cm
\ndt In this article, we describe how a series of three concrete steps leads us to uncover such an enhanced symmetry in $\mathcal N=8$ supergravity. We begin by noting that this theory is one in a series of maximal supergravity theories. These theories, in different spacetime dimensions, are descendants of the eleven-dimensional $\mathcal N=1$ supergravity. 
\vskip 0.1cm
\[ (\mathcal N=1, d=11)\ \text{ supergravity}\qquad \qquad  \]
\[ \qquad \xdownarrow{0.4cm} \qquad \qquad \qquad \]
\[ \quad \qquad \qquad (\mathcal N=8, d=5)\ \text{supergravity} \quad \rightarrow \quad E_{6(6)}\qquad \qquad \]
\[ \quad \qquad \qquad {(\mathcal N=8, d=4)\ \text{ supergravity} \quad \rightarrow \quad E_{7(7)}} \qquad  \qquad\]
\[ \quad \qquad \qquad{(\mathcal N=16, d=3)\ \text{ supergravity} \quad \rightarrow \quad E_{8(8)}}\qquad  \qquad\]
\[  \qquad  \xdownarrow{0.25cm} \qquad \qquad \qquad \]
\[ \qquad \qquad \qquad{(\mathcal N=32, d=1)\ \text{ supergravity} \quad \rightarrow  \quad E_{10}/ E_{11}\ }\qquad\] 
\vskip 0.3cm
\ndt Recent studies suggest that these symmetries extend further to infinite-dimensional groups~\cite{HN} and could perhaps answer questions regarding the origins of space itself. 
\vskip 0.2cm
\ndt Importantly, these dimensionally reduced supersymmetric theories retain considerable information regarding their higher dimensional parent. We used this recently~\cite{ABM}, to prove that ``oxidizing" the action of $\mathcal N=8$ supergravity to eleven dimensions, allows us to see signs of the $E_{7(7)}$ symmetry in $(\mathcal N=1, d=11)$ supergravity. 
\vskip 0.3cm
\ndt This is the key idea: realize a larger symmetry in the parent theory which was originally present only in the lower dimensional one. In particular, we demonstrate how the $E_{7(7)}$ symmetry in $(\mathcal N=8, d=4)$ supergravity may be enhanced to an $E_{8(8)}$ symmetry {\it {also in four dimensions}}. We will achieve this by first dimensionally reducing our theory to three dimensions and relating it by a field redefinition to the $E_{8(8)}$-invariant $(\mathcal N=16, d=3)$ supergravity. We will then ``oxidize" {\it {this}} theory to four dimensions in a careful manner, preserving the $E_{8(8)}$ symmetry that was picked up in $d=3$. 
\vskip 0.3cm
\ndt Here is a schematic of our plan of action.
\vskip 0.8cm

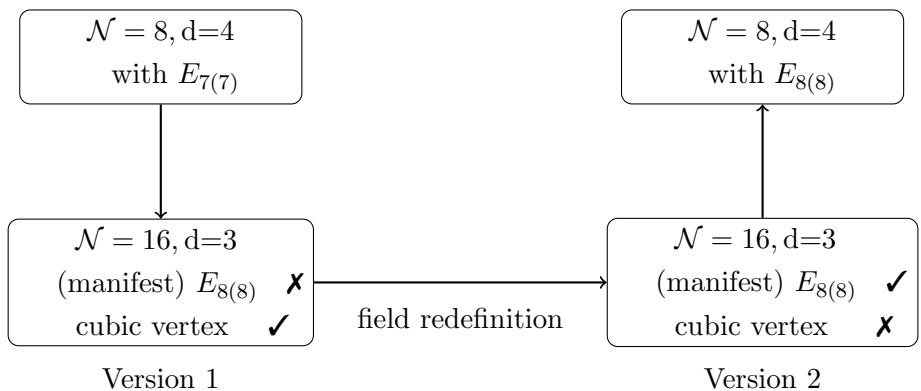
\begin{center}
\begin{tikzpicture}
\node (d4a) [forces] {\begin{varwidth}{14em} $ \qquad   \mathcal{N}=8,$\,d=4$ \qquad$   \vskip 0.1cm $\ \,  \qquad \ $with $ E_{7(7)}$  \end{varwidth}};
\node (d4b) [forces, xshift=8cm] {\begin{varwidth}{14em}$ \qquad \mathcal{N}=8,$\,d=4$ \qquad $  \vskip 0.1cm $ \ \qquad$  with $ E_{8(8)}$  \end{varwidth}} ;
\node (d3a) [forces, below of = d4a, yshift=-2.0cm] {\begin{varwidth}{14em}$\qquad \mathcal{N}=16, $\,d=3$ \qquad $  \vskip 0.1cm $\quad $ (manifest) $E_{8(8)}\quad $\ding{55}\vskip 0.1cm  $\quad \quad $cubic vertex \quad \ding{51}\quad    \end{varwidth}}node[midway, below, yshift=-4.0cm]{Version 1};
\node (d3b) [forces, xshift = 8cm, yshift=-3.0cm] {\begin{varwidth}{14em}$\qquad \mathcal{N}=16, $\,d=3$ \qquad $  \vskip 0.1cm $\quad $ (manifest) $E_{8(8)}\quad $\ding{51}\vskip 0.1cm  $\quad \quad $cubic vertex $\quad$ \ding{55}\quad    \end{varwidth}}node[midway, below, xshift= 8cm,yshift=-4.0cm]{Version 2};
\draw [->, line width=0.8pt , black] (d4a) -- (d3a);
\draw [->, line width=0.8pt, black] (d3b)-- (d4b);
\draw [->,line width=0.8pt, black] (d3a) --  (d3b) node[midway, below, yshift= -0.2cm]{\textcolor{black}{field redefinition}};
\end{tikzpicture}
\end{center}

\vskip 0.3cm
\ndt We work with light-cone coordinates, choosing $x^+=\frac{1}{\sqrt 2} (x^0+x^3)$ to be our time direction. Light-cone gauge is used for all fields (for example, we set $A^+=0$ for all vector fields) and the resulting constraints serve to eliminate unphysical degrees of freedom.
\vskip .3cm
\ndt {\it {$\mathcal N=8$ supergravity}}
\vskip 0.3cm
\ndt The $\mathcal N=8$ supergravity multiplet is comprised of one graviton, eight gravitinos, twenty-eight vectors, fifty-six spin one-half fermions and seventy real scalars. In light-cone superspace, spanned by eight Grassmann variables $\theta^m$ and their complex conjugates $\bar \theta_m\ (m=1, \ldots , 8)$, these 256 physical degrees of freedom can be captured in one constrained chiral superfield~\cite{LB3}

\bea
\label{superfield}
\begin{split}
\phi\,(\,y\,)\,=&\,\frac{1}{{\delp}^2}\,h\,(y)\,+\,i\,\theta^m\,\frac{1}{{\delp}^2}\,{\bar \psi}_m\,(y)\,+\,\frac{i}{2}\,\theta^m\,\theta^n\,\frac{1}{\delp}\,{\bar A}_{mn}\,(y)\ , \\
\,&+\,\ldots\, +\, \frac{1}{7!}\,(\theta)^7\,\delp\,\psi^u\,(y)\ +\,\frac{4}{8!}\,(\theta)^8\,{\delp}^2\,{\bar h}\,(y)\ ,
\end{split}
\eea
\newpage
\ndt In terms of this superfield, the action for $\mathcal N=8$ supergravity reads
\bea
\label{d=4 L}
\mathcal S \sim\int d^4 x\  d^8 \theta\  d^8 \bar \theta\;\; \Big\{&&\!\!\!\!\!\!\!\!\!- \bar{\phi}\ \frac{\Box}{\delp^4}\ \phi \nn \\
&&\!\!\!\!\!\!\!\!\!+ \frac{4}{3}\ \kappa\, \Big( \frac{1}{\delp^4} \bar \phi\ {\bar \partial}^2 \phi\ \delp^2 \phi-  \frac{1}{\delp^4} \bar \phi\ \delp \bar \del \phi\ \delp \bar \del \phi\Big ) + c.c. \Big\}
\eea
\vskip 0.2cm
\ndt where higher order terms in $\kappa$ are not shown. In this formalism, the $E_{7(7)}$ symmetry, which is a duality symmetry of the vector fields and a non-linear $\sigma$-model symmetry of the scalar fields in the covariant formalism, transforms all the physical fields in the supermultiplet.The 70 non-linear $\E/SU(8)$ coset transformations in light-cone superspace are given by

\be\label{E}
\delta \phi \ =\ -\frac{2}{\kappa}\,\theta^{klmn}_{}\,\overline\Xi^{}_{klmn}\ +\ \mathcal{O}(\kappa)\ +\cdots \nn
\ee

\ndt
where the order $\kappa$ terms are known~\cite{LB4} and $\overline\Xi^{}_{klmn}$ are transformation parameters. These coset transformations along with the linear $SU(8)$ transformations constitute the entire $E_{7(7)}$ algebra. The supermultiplet is therefore a representation of both the superPoincar\'e algebra and the $E_{7(7)}$, leading us to question which is more basic~\cite{ABM}.
\vskip 0.5cm
\ndt {\it Dimensional reduction to $d=3$}
\vskip 0.3cm
\ndt We begin by implementing the first arrow in the schematic. For this, we simply eliminate one of the spatial directions by setting $\bar\del=\del$ in (\ref {d=4 L}). The resulting Lagrangian, in $d=3$ (referred to as Version 1 in the picture below), has cubic interaction terms and does not yet exhibit the $E_{8(8)}$ symmetry, normally associated with maximal supergravity in $d=3$.
\vskip 1cm

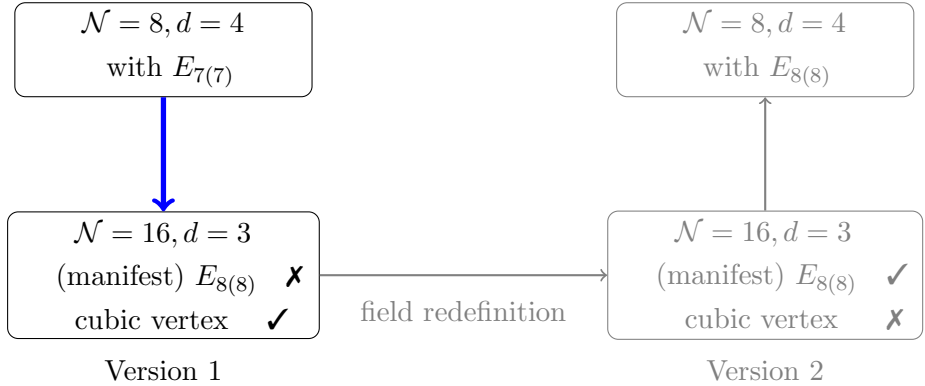
\begin{center}
\begin{tikzpicture}
\node (d4a) [forces] {\begin{varwidth}{14em} $ \qquad   \mathcal{N}=8,d=4 \qquad$   \vskip 0.1cm $\ \,  \qquad \ $with $ E_{7(7)}$  \end{varwidth}};
\node (d4b) [theory, xshift=8cm] {\begin{varwidth}{14em}$ \qquad\textcolor{gray}{ \mathcal{N}=8,d=4} \qquad $  \vskip 0.1cm  $ \ \qquad$ \textcolor{gray}{with $ E_{8(8)} $}  \end{varwidth}} ;
\node (d3a) [forces, below of = d4a, yshift=-2.0cm] {\begin{varwidth}{14em}$\qquad \mathcal{N}=16, d=3 \qquad $  \vskip 0.1cm $\quad $ (manifest) $E_{8(8)}\quad $\ding{55}\vskip 0.1cm  $\quad \quad $cubic vertex \quad \ding{51}\quad    \end{varwidth}}node[midway, below, yshift=-4.0cm]{Version 1};
\node (d3b) [theory, xshift = 8cm, yshift=-3.0cm] {\begin{varwidth}{14em}$\qquad \textcolor{gray}{\mathcal{N}=16, d=3} \qquad $  \vskip 0.1cm $\quad $ \textcolor{gray}{(manifest) $E_{8(8)}\quad $\ding{51}}\vskip 0.1cm  $\quad \quad $ \textcolor{gray}{cubic vertex $\quad$ \ding{55}} \quad    \end{varwidth}}node[midway, below, xshift= 8cm,yshift=-4.0cm]{\textcolor{gray}{Version 2}};
\draw [->, line width=2.0pt , blue] (d4a) -- (d3a);
\draw [->, line width=0.8pt, gray] (d3b)-- (d4b);
\draw [->,line width=0.8pt, gray] (d3a) --  (d3b) node[midway, below, yshift= -0.2cm]{\textcolor{gray}{field redefinition}};
\end{tikzpicture}
\end{center}

\vskip 0.5cm
\ndt {\it An aside: $E_{8(8)}$ invariant $\mathcal N=16, d=3$ supergravity}
\vskip 0.3cm
\ndt As a brief interlude, we discuss another $d=3$ theory that we refer to as Version 2. After the discovery of the $E_{7(7)}$ symmetry in $\mathcal N=8$ supergravity in four dimensions, it was conjectured that there should exist a maximal supergravity theory in three dimensions with an $E_{8(8)}$ symmetry. Such a theory with 128 scalars and 128 spin one-half fermions was later constructed~\cite{JS} accompanied by the important observation that its Lagrangian cannot involve interaction vertices of odd order (cubic, quintic and so on) -- since the 128 bosons and 128 fermions transform as two inequivalent spinor representations of the $SO(16)$ R-symmetry group. 
\vskip 0.3cm
\ndt The Lagrangian for this $SO(16)$- and $E_{8(8)}$- invariant theory in $d=3$ reads~\cite{LB5}
\bea
&\mathcal L &\sim -\ \bar \phi \frac{\Box}{\delp^4} \phi\ +\ \mathcal{O}(\kappa^2)\ . \nn
\eea
\vskip 0.2cm
\ndt
with the $E_{8(8)}$ symmetry being related to the embedding
\be
E_{8(8)} \supset SO(16)\ , \quad {\bf 248}\ =\ {\bf 120} + {\bf 128} \ .\nn
\ee
\vskip 0.1cm
\ndt The point here is that the $E_{8(8)}$ symmetry is not compatible with cubic interaction vertices.
\vskip 1.0cm
\ndt {\it {Relating the two versions of $d=3$ supergravity}}
\vskip 0.3cm
\ndt We are now ready to proceed to the next arrow in our schematic.
\vskip 0.5cm
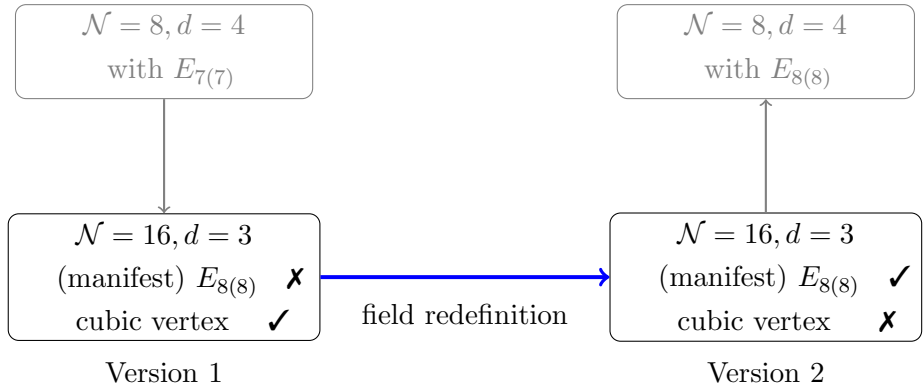
\begin{center}
\begin{tikzpicture}
\node (d4a) [theory] {\begin{varwidth}{14em} $ \qquad   \textcolor{gray}{\mathcal{N}=8,d=4} \qquad$   \vskip 0.1cm $\ \,  \qquad \ $\textcolor{gray}{with $ E_{7(7)}$ } \end{varwidth}};
\node (d4b) [theory, xshift=8cm] {\begin{varwidth}{14em}$ \qquad\textcolor{gray}{ \mathcal{N}=8,d=4} \qquad $  \vskip 0.1cm  $ \ \qquad$ \textcolor{gray}{with $ E_{8(8)} $}  \end{varwidth}} ;
\node (d3a) [forces, below of = d4a, yshift=-2.0cm] {\begin{varwidth}{14em}$\qquad \mathcal{N}=16, d=3 \qquad $  \vskip 0.1cm $\quad $ (manifest) $E_{8(8)}\quad $\ding{55}\vskip 0.1cm  $\quad \quad $cubic vertex \quad \ding{51}\quad    \end{varwidth}}node[midway, below, yshift=-4.0cm]{Version 1};
\node (d3b) [forces, xshift = 8cm, yshift=-3.0cm] {\begin{varwidth}{14em}$\qquad \mathcal{N}=16, d=3 \qquad $  \vskip 0.1cm $\quad $ (manifest) $E_{8(8)}\quad $\ding{51}\vskip 0.1cm  $\quad \quad $cubic vertex $\quad$ \ding{55}\quad    \end{varwidth}}node[midway, below, xshift= 8cm,yshift=-4.0cm]{Version 2};
\draw [->, line width=0.8pt , gray] (d4a) -- (d3a);
\draw [->, line width=0.8pt, gray] (d3b)-- (d4b);
\draw [->,line width=1.5pt, blue] (d3a) --  (d3b) node[midway, below, yshift= -0.2cm]{\textcolor{black}{field redefinition}};
\end{tikzpicture}
\end{center}
\ndt To relate the theory obtained by dimensional reduction of $(\mathcal N=8, d=4)$ supergravity (Version 1) to the $E_{8(8)}$- invariant theory (Version 2), cubic interaction terms must be eliminated from the former. This is achieved by the following field redefinition~\cite{ABM2}

\be
\label{redef}
\phi \rightarrow \phi\ =\ \phi'\ +\ \frac{1}{3}\ \kappa\ (\delp \phi'\ \delp \phi')\ +\ \frac{2}{3} \kappa\ \delp^4 \left( \frac{1}{\delp^3}\ \phi'\ \delp \bar \phi' \right)\ ,
\ee
\vskip 0.3cm
\ndt proving that the two versions are equivalent and exhibit an $E_{8(8)}$ symmetry.
\vskip 0.3cm
\ndt {\it {Oxidation to $d=4$}}
\vskip 0.3cm
\ndt Finally we implement the last arrow in our schematic (figure on next page). This involves starting with the $d=3$, $E_{8(8)}$- invariant theory and carefully oxidizing~\cite{ABR} it back to four dimensions. A new spatial derivative is introduced and the oxidation process respects both the $SO(16)$ and $E_{8(8)}$ symmetries. The exact dependence on the derivative is fixed by requiring that the resulting four-dimensional action be superPoincar\`e-invariant~\cite{ABM2}.

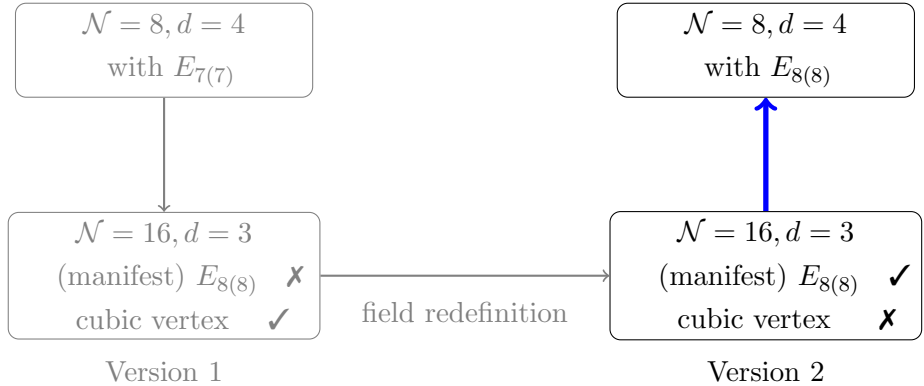
\begin{center}
\begin{tikzpicture}
\node (d4a) [theory] {\begin{varwidth}{14em} $ \qquad   \textcolor{gray}{\mathcal{N}=8,d=4 }\qquad$   \vskip 0.1cm $\ \,  \qquad \ $\textcolor{gray}{with $ E_{7(7)}$ } \end{varwidth}};
\node (d4b) [forces, xshift=8cm] {\begin{varwidth}{14em}$ \qquad \mathcal{N}=8,d=4 \qquad $  \vskip 0.1cm  $ \ \qquad$ with $ E_{8(8)}$  \end{varwidth}} ;
\node (d3a) [theory, below of = d4a, yshift=-2.0cm] {\begin{varwidth}{14em}$\qquad \textcolor{gray}{\mathcal{N}=16, d=3 }\qquad $  \vskip 0.1cm $\quad $ \textcolor{gray}{(manifest) $E_{8(8)}\quad $\ding{55}}\vskip 0.1cm  $\quad \quad $\textcolor{gray}{cubic vertex \quad \ding{51}}\quad    \end{varwidth}}node[midway, below, yshift=-4.0cm]{\textcolor{gray}{Version 1}};
\node (d3b) [forces, below of = d4b, yshift=-2.0cm] {\begin{varwidth}{14em}$\qquad \mathcal{N}=16, d=3 \qquad $  \vskip 0.1cm $\quad $ (manifest) $E_{8(8)}\quad $\ding{51}\vskip 0.1cm  $\quad \quad $cubic vertex $\quad$ \ding{55}\quad    \end{varwidth}}node[midway, below, xshift= 8cm,yshift=-4.0cm]{Version 2};
\draw [->, line width=0.8pt , gray] (d4a) -- (d3a);
\draw [->, line width=2.0pt, blue] (d3b)-- (d4b);
\draw [->,line width=0.8pt, gray] (d3a) --  (d3b) node[midway, below, yshift= -0.2cm]{\textcolor{gray}{field redefinition}};
\end{tikzpicture}
\end{center}
\vskip 0.3cm
\ndt This is our result: an action for maximal supergravity in four dimensions with the same field content as $\mathcal N=8$ supergravity, but with manifest $E_{8(8)}$ symmetry, at least upto second order in the coupling constant. Since maximal supergravity theories are unique, the two versions of $(\mathcal N=8,d=4)$ supergravity in the schematic must themselves be related by a field redefinition.
\vskip 0.3cm
\ndt This ``new" $E_{8(8)}$ symmetry, in four dimensions, is very likely to be behind many of the unexpected cancellations that are encountered. Note that in order to argue that the Hamiltonian is $E_{8(8)}$- invariant, we treat states as 128-dimensional spinors.  These are not the four-dimensional states of $(\mathcal N=8, d=4)$ supergravity. In order to argue that this symmetry is present in scattering amplitudes, in four dimensions, we must add up amplitudes such that the external states span the full 128-dimensional spinors. However, we can invoke supersymmetry which linearly relates the individual scattering amplitudes. 
\vskip 0.3cm
\ndt The obvious next step in this program is to incorporate this enhanced exceptional symmetry into the well established finiteness-analysis framework that exists for light-cone superspace~\cite{LB2,AKS1,AKS2}. Then we will be a step away from making an all-order statement regarding the finiteness of $\mathcal N=8$ supergravity, ending decades of speculation.

\end{document}